\documentclass[preprint,showkeys,showpacs,preprintnumbers,
                             amsmath,amssymb,aps,superscriptaddress,floatfix]{revtex4-1}
\usepackage{graphicx}
\usepackage[sort&compress]{natbib}
\usepackage{dcolumn}
\usepackage{bm}
\usepackage{latexsym,epsfig}
\usepackage{rotating}

\newcommand\lra[1]{\left\{#1\right\}}

\newcommand\sixj[1]{\lra{\begin{matrix}#1\end{matrix}}}

\begin{document}

\preprint{\today}

\title{Radium single-ion optical clock}

\author{O. O. Versolato}
\author{L. W. Wansbeek}
\author{K. Jungmann}
\author{R.~G.~E. Timmermans}
\author{L. Willmann}
\author{H. W. Wilschut}
\affiliation{KVI, University of Groningen, Zernikelaan 25,
                   9747 AA Groningen, The Netherlands \vspace{0.2cm}}
\date{\today}

\begin{abstract}
\noindent
We explore the potential of the electric quadrupole transitions $7s\,^2S_{1/2}$
- $6d\,^2D_{3/2}$, $6d\,^2D_{5/2}$ in radium isotopes as single-ion optical
frequency standards. The frequency shifts of the clock transitions due to
external fields and the corresponding uncertainties are calculated. Several
competitive $^A$Ra$^+$ candidates with $A=$ 223 - 229 are identified.
In particular, we show that the transition $7s\,^2S_{1/2}\,(F=2,m_F=0)$ - 
$6d\,^2D_{3/2}\,(F=0,m_F=0)$ at 828 nm in $^{223}$Ra$^+$, with no linear
Zeeman and electric quadrupole shifts, stands out as a relatively simple case,
which could be exploited as a compact, robust, and low-cost atomic clock
operating at a fractional frequency uncertainty of $10^{-17}$. With more
experimental effort, the $^{223,225,226}$Ra$^+$ clocks could be pushed
to a projected performance reaching the $10^{-18}$ level.
\end{abstract}

\pacs{} 
\keywords{}

\maketitle

\section{Introduction}
Optical atomic clocks based on ultranarrow optical transitions in single
laser-cooled trapped ions have demonstrated a stability and accuracy
significantly better than the $^{133}$Cs atom microwave frequency standard.
Transitions in various ions are presently under investigation as candidates
for optical frequency standards, including electric quadrupole transitions in
$^{40}$Ca$^+$ \cite{Mat08,Chw09},
$^{199}$Hg$^+$ \cite{Did01,Osk06,Sta07},
$^{88}$Sr$^+$ \cite{Mar04,Dub05},
and $^{171}$Yb$^+$ \cite{Sch05,Pei06},
hyperfine-induced electric dipole transitions in
$^{27}$Al$^+$ \cite{Ros07,Cho10a,Cho10b},
and $^{115}$In$^+$ \cite{Wan07}
and an electric octupole transition in
$^{171}$Yb$^+$ \cite{Hos09};
proposals also exist for
$^{137}$Ba$^+$ \cite{For05}
and $^{43}$Ca$^+$ \cite{Cha04}.
These ion clocks currently operate at  fractional frequency uncertainties
$\delta\nu/\nu$ ranging from $10^{-16}$ to below $10^{-17}$, with projected
accuracies reaching the $10^{-18}$ level. The ultimate performance of each
clock depends on the atomic structure of the ion, the sensitivity of the transition
to the external environment, and the complexity of the experimental setup
needed to operate the clock.

\begin{figure}[t]
\includegraphics[width = 15.0cm, angle = 0]{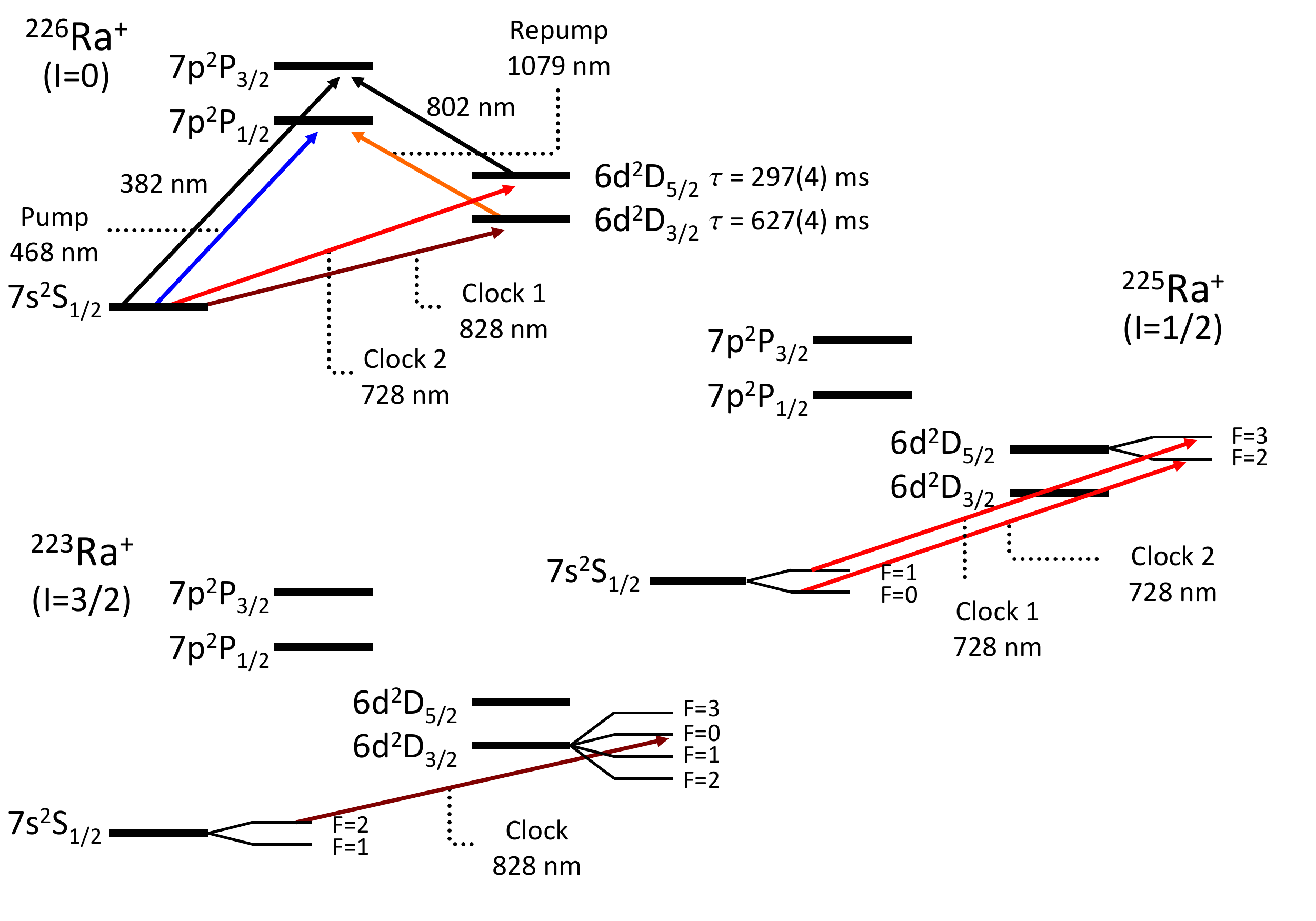}
\caption{(Color online) $^{223,225,226}$Ra$^+$ level scheme with wavelengths
taken from Ref.~\cite{Ras33} and lifetimes from Ref.~\cite{Sah07}. The clock
transitions are indicated; in $^{225}$Ra$^+$ and $^{226}$Ra$^+$ two clock
transitions are considered.}
\label{fig:levels}
\end{figure}

At our institute an experiment is in progress~\cite{Ver10} to measure atomic
parity violation in single Ra$^+$ ions~\cite{Wan08}. This experimental setup
can be adapted for an investigation of a single-ion Ra$^+$ clock. In this
paper we explore the feasibility of using the strongly forbidden
electric quadrupole transitions $7s\,^2S_{1/2}$ - $6d\,^2D_{3/2}$ at 828 nm
and $7s\,^2S_{1/2}\,$-$\,6d\,^2D_{5/2}$ at 728 nm in a single laser-cooled
and trapped Ra$^+$ ion as a stable and accurate frequency standard
\cite{Dzu00,Sah07,Sah09}. Our studies are based on the available experimental
information about the Ra$^+$ ion and on many-body atomic theory. The relevant
energy levels of $^{223,225,226}$Ra$^+$ and the proposed clock transitions
are shown in Fig.~\ref{fig:levels}. The $6d\,^2D_{3/2}$ and $6d\,^2D_{5/2}$
levels have a lifetime of 600 and 300 ms \cite{Sah07}, respectively, corresponding
to a $Q$ factor $\sim 10^{15}$ for the clock transitions. 

A major advantage of Ra$^{+}$ is that all the required wavelengths for cooling
and repumping and for the clock transition can easily be made with off-the-shelf
available semiconductor diode lasers, which makes the setup compact, robust,
and low-cost compared to clocks that operate in the ultraviolet. Moreover,
in odd radium isotopes clock transitions are available that are insensitive to
electric quadrupole shifts of the metastable $6d\,^2D_J$ levels. Such shifts are
an important limiting factor for several other ion clocks \cite{Mar09}. The radium
isotopes under consideration are mostly readily available from low-activity sources.

\setlength{\floatsep}{12pt}
\setlength{\textfloatsep}{12pt}

Optical clocks are important tools to test the fundamental theories of physics.
They are particularly useful in laboratory searches for possible spatial and
temporal variations of the physical constants that define these theories.
Such searches are strongly motivated by cosmological theories that unify
gravity and particle physics (see {\it e.g.} Ref. \cite{Bar10}). Laboratory tests
have placed strong limits on the temporal variation of the electron-to-proton
mass ratio $m_e/m_p$ \cite{Ros08,Bla08,Rei06} and the fine-structure
constant $\alpha$. The most stringent limit on the latter was obtained
by comparing two ultrasensitive ion clocks ($^{27}$Al$^+$ and
$^{199}$Hg$^+$) over the period of a year, yielding a limit
$\dot{\alpha}/\alpha = (-1.6\pm 2.3) \times 10^{-17}$/y \cite{Ros08}.
The sensitivity to $\dot{\alpha}/\alpha$ results from relativistic
contributions to the energy levels that are of order
${\cal O}$($Z^2\alpha^2$), favoring heavy atomic systems like
$^{199}$Hg$^+$. The Ra$^{+}$ clock transition has a comparably high
intrinsic sensitivity \cite{Dzu00, Sah07,Fla08} but of opposite sign
to that of $^{199}$Hg$^+$, making it a promising alternative
candidate for testing the time variation of $\alpha$. Ra$^+$ is also
very sensitive to variations in the quark masses \cite{Fla06,Din09}.

\section{Radium isotopes}
Radium offers a wide range of short- and long-lived isotopes with even and odd
nuclear spin that could be considered for use as optical frequency standards.
Only trace quantities of radium are needed to operate a single-ion Ra$^+$ clock,
but demands on the half-life and the ease of production limit the options. The
half-life of the isotope should be long compared to the excited $6d\,^2D_J$ 
level coherence time ($\sim$ seconds) required to address the ion with lasers.
Further, it is preferable from an experimental point of view to be able to trap the
ions for a longer time, at least a few minutes. 

\begin{table*}[t]
\caption{Long-lived neutron-rich isotopes of radium with their lifetime and nuclear
spin $I$~\cite{Aud03}, magnetic moments $\mu_I$ in units of $\mu_N$~\cite{Arn87}
and quadrupole moments $Q$ in barn~\cite{Neu89}. Also shown are the decay
series the isotopes occur in, and possible low-activity production sources; $A=227$
and 229 have to be produced by nuclear reactions.}
\begin{ruledtabular}
    \centering
    \begin{tabular}{ccccccc}
  $A$ &  Half-life & $I$ &$\mu_I$ &$Q$\footnotemark[1]&Decay series & Source \\ \hline
  223 &  11.43 d  &  3/2 &0.2705(19) &1.254(66)& $^{235}$U    & $^{227}$Ac (21.8 y)   \\
  224 &    3.66 d  &   0    &0&0&   $^{232}$Th  & $^{228}$Th (1.9 y)  \\
  225 &   14.9 d   &  1/2  &$-$0.7338(5) &0& $^{233}$U & $^{229}$Th (7.34 ky)  \\
  226 &   1.6 ky    &   0    &0&0&   $^{238}$U   & $^{226}$Ra, $^{230}$Th (75.4 ky)  \\
  227 &  42.2 m   &  3/2  & $-$0.4038(24)  &1.58(11)&          -           & - \\
  228 &    5.75 y  &   0     &0&0&  $^{232}$Th & $^{228}$Ra \\
  229 &    4.0 m   &  5/2  & 0.5025(27) &3.09(19)&            -          & -
 \end{tabular}
 \label{table:isotopes}
\end{ruledtabular}
\footnotetext[1]{The uncertainties were obtained by adding in
quadrature the uncertainties given in Ref.~\cite{Neu89}.}
\end{table*}

The light (neutron-poor) isotopes $A=$ 209-214, with half-lives that range
from several seconds up to a few minutes, have been produced at the KVI by
fusion-evaporation reactions~\cite{Ver10,Shi09}. A possible clock candidate
could be $A=213$, which has a half-life of 2.7 m; it is similar to the isotope
$A=225$, which we consider in detail below. We focus in this paper on the
heavier (neutron-rich) isotopes with $A =$ 223-229, because they have a
half-life of longer than one minute, and, moreover, most of them occur in
the decay series of uranium or thorium and therefore can be produced in
sufficient quantities with a low-activity source, so that no accelerator is required.
Table~\ref{table:isotopes} gives an overview of these isotopes, with their half-life,
nuclear spin, and possible production methods. The nuclear magnetic moments
and quadrupole moments listed are used to calculate the hyperfine constants
of the $6d\,^2D_{3/2}$ and $6d\,^2D_{5/2}$ levels of the odd isotopes for which
no experimental results are available.

For Ra$^+$ optical-clock purposes, the even isotopes $A=$ 224, 226, and 228,
with zero nuclear spin, are very similar and spectroscopically relatively simple.
They are analogous to the $^{40}$Ca$^+$ and $^{88}$Sr$^+$ clocks.
$^{226}$Ra and $^{228}$Ra are available as a source; $^{226}$Ra$^+$ can
also be taken from a $^{230}$Th source, in which case there is no need to ionize
the atoms. We limit ourselves to $^{226}$Ra$^+$, which is the most easily available
isotope, and we consider two transitions, namely $7s\,^2S_{1/2}$ - $6d\,^2D_{3/2}$
and $7s\,^2S_{1/2}$ - $6d\,^2D_{5/2}$, as indicated in Fig.~\ref{fig:levels}.

In the odd isotopes, with nonzero nuclear spin, the presence of hyperfine structure
gives two advantages. First, in all odd isotopes $m_F=0\leftrightarrow m_F'=0$
transitions exist, which are insensitive to the linear Zeeman shift. Moreover, the odd
isotopes offer several transitions between specific hyperfine levels that in first order
do not suffer from the Stark shift due to the electric quadrupole moment of the
$6d\,^2D_J$ level. In particular, we study the transition $7s\,^2S_{1/2}\,(F=2,m_F=0)$
- $6d\,^2D_{3/2}\,(F=0,m_F=0)$ in $^{223}$Ra$^+$ (no linear Zeeman and
quadrupole shifts) and $7s\,^2S_{1/2}\,(F=1,m_F=0)$ - $6d\,^2D_{5/2}\,(F=3,m_F=\pm 2)$
in $^{225}$Ra$^+$ (no quadrupole shift), see Fig.~\ref{fig:levels}. In addition, we
consider the transition $7s\,^2S_{1/2}\,(F=0,m_F=0)$ - $6d\,^2D_{5/2}\,(F=2,m_F= 0)$
in $^{225}$Ra$^+$ (no linear Zeeman shift), which resembles the $^{199}$Hg$^+$
clock. We also include the isotopes $A =227$ and 229, although their half-lives are
rather short and they must be produced in nuclear reactions. Specifically, we consider
the transitions $7s\,^2S_{1/2}\,(F=2,m_F=0)$ - $6d\,^2D_{3/2}\,(F=0,m_F=0)$ in
$^{227}$Ra$^+$ and $7s\,^2S_{1/2}\,(F=2,m_F=0)$ - $6d\,^2D_{5/2}\,(F=0,m_F=0)$
in $^{229}$Ra$^+$; both transitions are free from linear Zeeman and quadrupole
shifts.

\section{Sensitivity to external field shifts}
All proposed optical frequency standards are sensitive to external perturbations
due to the electric and magnetic fields present in the trap. These perturbations
cause unwanted systematic shifts of the frequency of the clock transition.
Although for a large part these shifts themselves can be corrected for, there is
a remaining uncertainty associated with each shift due to limited experimental
or theoretical accuracy. In this Section, we will investigate the sensitivity to the
external fields of the candidate Ra$^+$ clock transitions for the different isotopes.
Input for the required atomic-structure quantities is taken from the recent KVI
experiment \cite{Ver10} and from experiments at the ISOLDE facility at CERN
\cite{Arn87,Wen87,Neu89}. The wavelengths of the relevant transition are taken
from Ref. \cite{Ras33}. When no experimental data are available, we rely on
atomic many-body theory calculations. 

In the following, we briefly discuss the relevant shifts point-wise. The shift of the
clock transition is defined as the shift of the excited $6d\,^2D_J$ level minus
the shift of the $7s\,^2S_{1/2}$ ground state. The results of our
calculations for the different Ra$^+$ isotopes are summarized below and divided
into a {\em sensitivity}, see Table \ref{table:sensitivities}, and an {\em uncertainty},
see Table \ref{table:uncertainties}. The theoretical expressions for the various
external-field shifts can either be found in the literature or they are straightforward
to derive; for completeness, the most important ones are given. In the following, we
assume that one single laser-cooled radium ion is trapped in a radiofrequency
(RF) electric quadrupole field, {\em i.e.} in a Paul trap.

 \subsection{Doppler shifts}
The motion of an ion in a Paul trap can be described by a secular oscillation
with a superimposed micromotion oscillation \cite{Mah05}. The micromotion
oscillation is directly driven by the RF field applied to the trap. Any movement
of the ion in the trap can, via the Doppler effect, cause broadening and
shifts of the frequency of the clock transition. This effect is important even
when the ion is laser-cooled to the Doppler limit. In the Lamb-Dicke regime
\cite{Lei03}, which can be reached by Doppler cooling on the strong
$7s\,^2S_{1/2}\,$-$\,7p\,^2P_{1/2}$ transition at 468 nm, the oscillation amplitude
is small compared to the laser-light wavelength, and first-order Doppler shifts are
essentially negligible \cite{Liz07,Win79}. Second-order Doppler shifts are still
present. However, it can be shown that for a heavy ion like Ra$^+$ this shift
is negligible in the Doppler cooling limit \cite{Cha04}, with a projected fractional
frequency uncertainty in the low $10^{-19}$ levels. It is, of course, a major
challenge to achieve this limit experimentally \cite{Cho10a}; excess
micromotion of the ion, caused by electric fields that displace the ion from
the middle of the RF pseudopotential, needs to be minimized.

\begin{table*}[t]
\caption{The available experimental and theoretical hyperfine
structure constants (in MHz) of the $7s\,^2S_{1/2}$,
$6d\,^2D_{3/2}$, and $6d\,^2D_{5/2}$ levels of the relevant odd
isotopes of Ra$^+$. The values $A'_J$ for the isotopes for which
no data was available were calculated with $A'_J =
(I/I')\times(\mu'_I/\mu)A_J$, while for $B_J$ we used $B'_J = (Q'/Q)
B_J$. The reference values are printed bold. For the $7\,s^2S_{1/2}$
two different sets of experimental data were available; we used the
underlined values. The experimental value for $A_D$ of
the $6d\,^2D_{3/2}$ level of $^{213}$Ra was used to calculate $A_D$
for the $6d\,^2D_{3/2}$ levels of the heavy isotopes; the $^{213}$Ra
magnetic moment  used is $\mu_I=0.6133(18)$ \cite{Arn87}. There are
no data for the $B_D$ coefficient of the $6d\,^2D_{3/2}$
level, nor for $A_D$ and $B_D$ of the $6d\,^2D_{5/2}$ level.
Consequently, we used the theoretical values listed and estimated
the uncertainty of the $A_D$ coefficients of the $6d\,^2D_{5/2}$ to
be 3\%, and the uncertainty of all $B_D$ coefficients conservatively
as 10\%.}
\centering
\begin{tabular}{llcccccc}
\hline \hline
 &&&{$7s\,^2S_{1/2}$}&\multicolumn{2}{c}{$6d\,^2D_{3/2}$}
                                        &\multicolumn{2}{c}{$6d\,^2D_{5/2}$}\\
&&&$A_S$&$A_D$&$B_D$&$A_D$&$B_D$\\
\hline  $^{213}$Ra$^+$&Expt.& \cite{Wen87} &22920.0(6.0)&- &0 &- &0 \\
&Expt.&\cite{Ver10}&-&{\textbf{528(5)}} &0 &- &0\\
$^{223}$Ra$^+$& Expt.& \cite{Neu89}&\underline{3404.0(1.9)}&-&-&- &- \\
&Expt.& \cite{Wen87} &3398.3(2.9)&- &- &- &-\\
&Theory& \cite{Sah07}&3567.26&77.08&
\textbf{383.88}&$-$\textbf{23.90}&\textbf{477.09}\\
&Theory& \cite{Pal09}&{3450}&{79.56}&-&$-$24.08&-\\
$^{225}$Ra$^+$&Expt.& \cite{Neu89}&\underline{$-$27731(13)}&-&0 &- &0\\
&Expt.& \cite{Wen87} &$-$27684(13)&-&0 &- &0\\
&Theory& \cite{Sah07}&$-$28977.76&$-$626.13&0&194.15&0\\
$^{227}$Ra$^+$ &Expt.& \cite{Wen87} &$-$5063.5(3.1)&- &- &-&-  \\
 $^{229}$Ra$^+$ &Expt.& \cite{Wen87} &3789.7(2.3)&- &- &-&- \\
 \hline\hline
\end{tabular}
\label{table:hyperfine}
\end{table*}

\subsection{Zeeman shifts}
Magnetic fields in the trap lead to frequency shifts of the clock transition via
the linear and quadratic Zeeman effect. For the transitions that suffer from the
linear Zeeman effect it is hard to quantify the theoretical uncertainty, because
the achievable accuracies depend on experimental details. In these cases,
multiple transitions $m_F\leftrightarrow m_F'$ can be used to average out
the linear effect to the desired level of accuracy.
The linear Zeeman shift is absent in $m_F=0\leftrightarrow m_F'=0$transitions,
in which case the quadratic Zeeman shift $\Delta\nu_\textrm{\tiny QZ}$ becomes
the dominant source of uncertainty. For the state $|\gamma,I,J;F,m_F\rangle$
it is given by
\begin{eqnarray}
   h\Delta \nu_\textrm{\tiny QZ}(\gamma,I,J,F,m_F)&=&
       \left(g_J\mu_B-g_I\mu_N\right)^2B^2J(J+1)(2J+1)\times \nonumber \\
&&\sum_{F'}
      \left\{\begin{array}{ccc}J&F'&I\\F&J&1\end{array}\right\}^2\left(
     \begin{array}{ccc}F&1&F'\\-m_F&0&m_F\end{array}\right)^2\frac{(2F+1)(2F'+1)}
     {E-E^\prime} \ ,
\end{eqnarray}
where the magnetic field $B$ is taken along the $z$-axis; $\gamma$ labels
all quantum numbers that are not specified.
We consider only couplings to the hyperfine-structure partners, since other
contributions will be suppressed; therefore, the quadratic Zeeman effect is
negligible in the even isotopes. The Zeeman shifts can be calculated from the
hyperfine structure constants $A_{S,D}$ and $B_D$ of the $7s\,^2S_{1/2}$,
$6d\,^2D_{3/2}$, and $6d\,^2D_{5/2}$ levels, and the electron and nuclear
$g$-factors. Table \ref{table:hyperfine} lists the available experimental and
theoretical values of $A_{S,D}$ and $B_D$ of the relevant odd isotopes.

\subsubsection{DC Zeeman shift}
DC Zeeman shifts are caused by the static applied magnetic field
present in the trap. We assume a magnetic field of 1 mG, which is a
typical value needed to split the Zeeman degeneracies to order
$\sim$10 kHz needed for proper state addressing. Passive shielding
of an ion trap against magnetic fields has achieved $\leq 10$ $\mu$G
field stability~\cite{She05}. This experimental number is taken as the
uncertainty in the magnetic field strength in Table~\ref{table:uncertainties}.
In order to calculate the uncertainty in the resulting shifts, the uncertainties
in $A_D$ and $B_D$, in the magnetic field ($\sim 10$ $\mu$G), and in
the $g_J$ values were taken into account. For $g_J$ the free-electron
values were used with a conservative 1\% uncertainty. The uncertainties
due to $g_I$ and the parameters associated with the $7s\,^2S_{1/2}$
state are negligible.

\subsubsection{AC Zeeman shift}
The RF voltages applied to the trap electrodes require rather large currents to flow.
These currents give rise to an AC magnetic field in the trap center. In a perfect
geometry, when the currents to all electrodes are equal, the individual contributions
of the electrodes will cancel each other and the net magnetic field will be zero.
However, this cancellation could be far from complete \cite{Ros08}. The oscillating
magnetic field averages over the clock interrogation time (which is of the order of
the $6d\,^2D_J$-level lifetime), which is long compared to typical RF periods
(0.1-1 $\mu$s). Therefore, the expressions for the DC Zeeman effect can be used, 
with a rms magnetic field. For the $^{199}$Hg$^+$ clock this magnetic field is
conservatively estimated to be of the order $\sim$ mG \cite{Ros08}. We use 1 mG
as estimate in Table \ref{table:uncertainties}, because for Ra$^+$ the mass and
other trapping parameters  are similar. The resulting AC Zeeman shift proves to be
one of the largest shifts. Therefore, it is important to work with a rather weak trap
potential, as the average magnetic field scales with RF power. By varying the trap
parameters the AC Zeeman shift can be measured. Moreover, averaging schemes
that exploit the hyperfine structure could significantly reduce the uncertainty in the
AC Zeeman shift. In this way it should be possible to reduce this uncertainty to the
level of 25\% of the shift itself; this is the uncertainty used in Tabel \ref{table:uncertainties}.

\subsection{Stark shifts}
Stark shifts result both from static electric fields (causing DC Stark shifts) and
from dynamic electric fields (causing AC Stark or light shifts). First, quadratic
dipole Stark shifts are discussed, which are caused by the interaction of the
dipole moment of the atom with the electric field.  Next, we discuss quadrupole
Stark shifts, caused by the interaction of the quadrupole moment of the atom
with the gradient of the electric trap field; we look at both linear and quadratic
quadrupole Stark shifts.

\begin{table}[t]
\caption{Dipole scalar, $\alpha_0^1$, and tensor, $\alpha_2^1$, polarizabilities,
in units of $4\pi\varepsilon_0a_0^3$, and quadrupole moments, $\Theta$,
in units of $ea_0^2$, for the $7s\,^2S_{1/2}$, $6d\,^2D_{3/2}$, and 
$6d\,^2D_{5/2}$ levels in Ra$^+$.}
\begin{ruledtabular}
\centering
\begin{tabular}{cccrr}
& Ref. &  $7s\,^2S_{1/2}$&$6d\,^2D_{3/2}$&$6d\,^2D_{5/2}$\\
\hline
$\alpha_0^1$ & \cite{Sah09}& 104.54(1.5)&83.71(77)      &82.38(70)\\
                          & \cite{Pal09} & 106.22         &                       &                  \\
$\alpha_2^1$ & \cite{Sah09}&      -               &$-$50.23(43)&$-$52.60(45)\\
$\Theta$          & \cite{Sah07}&      -               & 2.90(2)  & 4.45(9) \\
\end{tabular}
\end{ruledtabular}
\label{table:pols}
\end{table}

\subsubsection{DC dipole Stark shift}
The theory of the static quadratic dipole Stark shift was developed by Angel and
Sandars \cite{Ang68}. For the state $|\gamma;J,m_J\rangle$ this shift is given by
\begin{equation}
h\Delta \nu_\textrm{\tiny DCDS}(\gamma,J,m_J) =
   -\frac{1}{2}\alpha_0^1(\gamma,J)E^2-\frac{1}{2}\alpha_2^1(\gamma,J)
    \frac{3m_J^2-J(J+1)}{2J(2J-1)}(3E_z^2-E^2) \ ,
\end{equation}
where $E$ is the DC electric field strength, $\alpha_0^1$ and $\alpha_2^1$
are the scalar and tensor polarizabilities, respectively. In Table \ref{table:pols}
the available theoretical calculations for these polarizabilities are listed for
the $7s\,^2S_{1/2}$, $6d\,^2D_{3/2}$, and $6d\,^2D_{5/2}$ levels in Ra$^+$;
we used the results of Ref.~\cite{Sah09} in our calculations. The polarizabilities
for the hyperfine levels $|\gamma,I,J;F,m_F\rangle$ are calculated using
\begin{equation}
   \alpha_k^1(\gamma,I,J,F) = (-1)^{J +I+F+k}(2F+1)
   \sixj{F&F&k\\J &J&I}\alpha_k^1(\gamma,J) \ .
\end{equation}
For an ion laser-cooled to the Lamb-Dicke regime, DC electric fields at the position
of the ion can be reduced to $< 10$ V/m in the process of minimizing the micromotion
\cite{Ros08}. This is the field uncertainty that we assume to estimate the fractional
uncertainty in Table \ref{table:uncertainties} in a worst case scenario, {\em i.e.}
$E_z=E$.

The main source of DC dipole Stark shifts, however, is the presence of black-body
(BB) radiation due to the nonzero temperature $T$ of the trap and its surroundings.
The energy shift of a level with dipole scalar polarizability $\alpha^{1}_0$ in a
BB electric field is given by \cite{Aro07}
\begin{equation}
   h\Delta\nu_{\text{BB}}(\gamma,J,m_J) = -\frac{1}{2}\left(8.319\,{\rm V/cm}\right)^2	
                       \bigg(\frac{T({\rm K})}{300}\bigg)^4\alpha^1_0(\gamma,J)(1+\eta) \ ,
\end{equation}
where $\eta$ is a small calculable term associated with dynamical corrections; it is
of the order of a few percent \cite{Por06} and therefore it can be neglected compared
to the overall 10\% uncertainty given in Table \ref{table:sensitivities}, which is mainly
due to the theoretical uncertainties in the polarizabilities. The BB radiation is assumed
to be isotropic, so the tensor polarizability plays no role. Since the BB radiation
shift results in a relatively large fractional frequency uncertainty at room temperature
$T = 293$ K (see Table \ref{table:uncertainties}), the calculation was also performed
for liquid-nitrogen temperature, $T = 77$ K (the $^{199}$Hg$^+$ clock operates at 4 K).
We assume an uncertainty in the temperature of 1 K, as in Ref. \cite{Lud08}.

\begin{sidewaystable}[t]
\centering \caption{Overview of the sensitivities to external-field shifts
with the associated uncertainties between brackets: linear (LZ) and
quadratic (QZ) Zeeman, dipole Stark (DS) DC and AC, and linear or
quadratic quadrupole Stark (QS). The quoted uncertainties are derived
from a Monte Carlo model, taking into account the uncertainties for all
parameters as quoted in the text and previous Tables; $t\equiv(3E_z^2-E^2)/(2E^2)$
parametrizes the tensor part of the DC dipole Stark shift.} 
\centering
\begin{ruledtabular}
\begin{tabular}{llllllll}
    Isotope & Transition & LZ & QZ  & DS DC & DS AC & QS \\
     &  & & mHz/mG$^2$ &mHz V$^{-2}$ cm$^{2}$ & mHz $\mu$W mm$^{-2}$
     & mHz V$^{-1}$ cm$^{2}$ \\ \hline
  $^{223}$Ra$^+$ & $7s\,^2S_{1/2}^{F=2,m_F=0}$\,-\,$6d\,^2D_{3/2}^{F=0,m_F=0}$
     &no &$4.9(7)$& 2.6(2) &$0.72(4)$ & $15(2) \times 10^{-9}$ \footnotemark[1]\\
  $^{225}$Ra$^+$(1)& $7s\,^2S_{1/2}^{F=1,m_F=0}$\,-\,$6d\,^2D_{5/2}^{F=3,m_F= \pm 2}$
     & yes&$0.75(3)$& 2.8(2) &$1.6(3)$ &$6.2(3) \times 10^{-9}$ \footnotemark[1] \\
  $^{225}$Ra$^+$(2)& $7s\,^2S_{1/2}^{F=0,m_F=0}$\,-\,$6d\,^2D_{5/2}^{F=2,m_F= 0}$
     &no &$-1.28(5)$& $2.8(2) - 5.23(5) t$ &$1.2(3)$&24.1(5) \\
  $^{226}$Ra$^+$(1)&
     $7s\,^2S_{1/2}^{m_J=\pm\frac{1}{2}}$\,-\,$6d\,^2D_{3/2}^{m_J=\pm\frac{3}{2}}$
     &yes&0 & $2.6(2) + 6.25(5) t$  & $0.9(2)$ & $-19.6(1)$ \\
  $^{226}$Ra$^+$(2)&
     $7s\,^2S_{1/2}^{m_J=\pm\frac{1}{2}}$\,-\,$6d\,^2D_{5/2}^{m_J=\pm\frac{3}{2}}$
     &yes&0 & $2.8(2) - 1.30(1)t$ & $1.5(4)$ & $6.0(1)$ \\
  $^{227}$Ra$^+$& $7s\,^2S_{1/2}^{F=2,m_F=0}$\,-\,$6d\,^2D_{3/2}^{F=0,m_F=0}$
     &no &$2.8(2)$& 2.6(2) &0.72(4)
    &  $5.9(4) \times 10^{-9}$ \footnotemark[1] \\
  $^{229}$Ra$^+$& $7s\,^2S_{1/2}^{F=2,m_F=0}$\,-\,$6d\,^2D_{5/2}^{F=0,m_F=0}$
     & no & $27(3)$ & $2.8(2)$ &$1.6(3)$& $12(1) \times 10^{-9}$ \footnotemark[1]
      \\ \hline
  $^{43}$Ca$^+$ & $4s\,^2S_{1/2}^{F=4,m_F=0}$\,-\,$3d\,^2D_{5/2}^{F=6,m_F= 0}$
    &no&90.5 \cite{Cha04} &$5.6(4)+2.1(2) t$ \cite{Cha04}&8(8) \cite{Cha04}
    &8.1 \cite{Cha04}\\
  $^{199}$Hg$^+$&
    $5d^{10}6s\,^2S_{1/2}^{F=0,m_F=0}$\,-\,$5d^96s^2\,^2D_{5/2}^{F=2,m_F= 0}$
    &no&0.18925(28) \cite{Ita00}&$-$1.14 \cite{Ita00}&-\footnotemark[2]&$-3.6$ \cite{Ita00} \\
  $^{88}$Sr$^+$& $5s\,^2S_{1/2}^{m_J=\pm 1/2}$\,-\,$4d\,^2D_{5/2}^{m_J=\pm 5/2}$
    &yes&0&$4.6(2)$ \cite{Mar09}
    &$-2.24$ \cite{Mad04} &$-18(2)$ \cite{Bar04}
\end{tabular}
\end{ruledtabular}
\label{table:sensitivities}
\footnotetext[1]
{These are second-order quadrupole shifts with units mHz (V$^{-1}$ cm$^{2}$)$^2$.}
\footnotetext[2]
{The uncertainty caused by the AC Stark shift was measured to contribute less than
                   $2\times 10^{-17}$ to the fractional frequency uncertainty \cite{Osk06}.}
\end{sidewaystable}

\begin{sidewaystable}[h]
\centering
\caption{Overview of the shifts, in mHz, due to the external fields with the
associated uncertainties between brackets. The values and uncertainties
are derived from those in Table \ref{table:sensitivities}, taking into account
the field uncertainties as explained in the text. At the bottom of the Table
the resulting fractional frequency uncertainties $\delta \nu/\nu$ caused by
the external-field shifts are given for different scenarios; $\delta \nu$ indicates
the uncertainty in a certain shift, rather than the shift itself. The transitions for
the different isotopes are as given in Table \ref{table:sensitivities}.}
\centering
\begin{ruledtabular}
\begin{tabular}{llllllllll}
    Shift&$^{223}$Ra$^+$&$^{225}$Ra$^+$(1)&$^{225}$Ra$^+$(2)
    &$^{226}$Ra$^+$(1)&$^{226}$Ra$^+$(2)&$^{227}$Ra$^+$&$^{229}$Ra$^+$&\\
    \hline
    LZ &no&yes&no&yes&yes&no&no&\\
    QZ, DC &4.9(7)&0.74(3)&$-1.28(5)$&0&0&2.8(2)&$27(3)$&\\
    QZ, AC &(1.2)&(0.19)&($-0.32$)&0&0&(0.7)&(6.8)&\\
    BB, 293(1) K &163(14)&173(13)&173(13)&163(13)&174(13)&163(13)&174(13)&\\
    BB, 77(1) K &0.78(8)&0.83(8)&0.83(8)&0.78(8)&0.83(7)&0.78(7)&0.83(8)&\\
    DS, DC Scalar&(0.026)&(0.028)&(0.028)&(0.026)&(0.028)&(0.026)&(0.028)& \\
    DS, DC Tensor&0&0&($-0.05$)&$(-0.06)$&($-0.013$)&0&0& \\
    DS, AC &$0.72(4)$&$1.6(3)$&$1.2(3)$&$0.9(2)$&$1.5(4)$&$0.72(4)$&$1.6(3)$&\\
    QS &$1.5(2)$&$0.62(3) $&$(24.1) \times 10^{3} $&
                 $(-19.6) \times 10^{3}$&$(6.0) \times 10^{3}$&$0.59(4)$&$1.2(1)$&\\
    \hline
    Total shift (293 K) &170(14)&177(13)&$(24)\times 10^{3}$&$(20)\times 10^{3}$  
    &$(6.0)\times 10^{3}$&$167(14)$&$203(15)$\\
    Total shift (77 K) &7.9(1.4)&3.8(4)&$(24)\times 10^{3}$&$(20)\times 10^{3}$    
    &$(6.0)\times 10^{3}$&$4.9(7)$&$30(7)$&\\
    Total shift (293 K, no QS) &&&173(13)&164(13)&175(13)&&\\
    Total shift (77 K, no QS) &&&0.7(4)&1.5(2)&2.3(4)&&&\\
    \hline
   Total $\delta\nu/\nu$ (293 K)&$3.7\times 10^{-17}$&$3.2\times 10^{-17}$    
   &$5.9\times 10^{-14}$&$5.4\times 10^{-14}$&$1.5\times 10^{-14}$    
   &$3.7\times 10^{-17}$&$3.6\times 10^{-17}$&\\
    Total $\delta\nu/\nu$ (77 K)&$4.0\times 10^{-18}$&$9.1\times 10^{-19}$    
    &$5.9\times 10^{-14}$&$5.4\times 10^{-14}$&$1.5\times 10^{-14}$    
    &$2.1\times 10^{-18}$&$1.7\times 10^{-17}$&\\ 
    Total $\delta\nu/\nu$ (293 K, no QS)&&   
    &$3.2\times 10^{-17}$&$3.7\times 10^{-17}$&$3.3\times 10^{-17}$&&&\\
    Total $\delta\nu/\nu$ (77 K, no QS)&    
    &&$1.1\times 10^{-18}$&$4.9\times 10^{-19}$&$9.1\times 10^{-19}$&&&\\
\end{tabular}
\end{ruledtabular}
\label{table:uncertainties}
\end{sidewaystable}

\subsubsection{AC dipole Stark shift}
The most important cause of AC dipole Stark shifts is the laser locked to either
the 728 nm or 828 nm clock transition, since we assume that the cooling and
probing lasers are fully extinguished at the time of measurement. When the
laser propagates along the $z$-axis, the AC dipole Stark shift of a state
$|\gamma,J,m_J\rangle$ is given by \cite{Ros09}
\begin{equation}
   h\Delta\nu_\textrm{\tiny ACDS}(\gamma,J,m_J;\nu_L) = -
   \frac{I_L}{2\varepsilon_0c}\left(\alpha_0^1(\nu_L) +A\,\alpha_1^1(\nu_L)
   \frac{m_J}{2J}-\alpha_2^1(\nu_L)\frac{3m_J^2-J(J+1)}{2J(2J-1)}\right) \ ,
\end{equation}
where  $I_L$ is the intensity of the laser which we take as 1 $\mu$W/mm$^2$,
$\nu_L$ is its frequency at the clock transition, and $A$ is a numerical factor
whose value depends on the type of polarization. Further, $\alpha_0^1(\nu_L)$,
$\alpha_1^1(\nu_L)$, and $\alpha_2^1(\nu_L)$ are the dynamic scalar, vector,
and tensor polarizability, respectively, of the state $|\gamma,J,m_J\rangle$.
We choose the polarization such that $A = 0$, therefore we only need the
scalar and tensor polarizabilities. These are given by
\begin{eqnarray}
   \alpha_0^1(\gamma,J;\nu_L) & = &
   -\frac{2}{3(2J+1)}\sum_{\gamma'J'}|\langle \gamma'J'||D||\gamma
   J\rangle|^2 \frac{ \Delta E }{\left(\Delta E\right)^2-(h\nu_L)^2} \ , \\
   \alpha_2^1(\gamma,J;\nu_L) &=&-4 \sqrt{\frac{5}{6}}
   \left(\frac{J(2J-1)}{(2J+3)(J+1)(2J+1)}\right)^{1/2}(-1)^{2J} \times \nonumber \\
   && \sum_{\gamma'J'}(-1)^{J-J'}\sixj{1&1&2\\J&J&J'}
         |\langle\gamma'J'||D||\gamma J\rangle|^2
         \frac{\Delta E}{\left(\Delta E\right)^2-(h\nu_L)^2} \ ,
\end{eqnarray}
with $\Delta E = E-E^\prime$ and $D$ the dipole operator. For
$\nu_L\rightarrow 0$, the above equations reduce to their static counterparts.
In calculating the dynamic polarizabilities we use the values for the dipole matrix
elements given in Refs. \cite{Wan09,Pal09,Saf07}. In using this sum over the valence
states approach, we do not take the core contributions, which are of order 10\%
\cite{Sah07}, into account. However, the core contributions cancel since we look
at differential shifts, and these contributions are common. The remaining uncertainty
is due to neglected higher-order valence and valence-core couplings, and the
uncertainty in the dipole matrix elements.

\subsubsection{Quadrupole Stark shift}
The interaction of the atomic quadrupole moment with the gradient of an electric
field gives rise to an electric quadrupole shift. This shift is troublesome in several
optical frequency standards \cite{Mar09}. The expression used for the linear
quadrupole Stark shift is \cite{Ita00}
\begin{eqnarray}
   h\Delta\nu_{\mbox{\tiny{LQS}}}(\gamma,I,J,F,m_F) & = &
               A_{\mbox{\tiny{DC}}}\Theta(\gamma,J)
              \frac{2\left[F(F+1)-3m_F^2\right](2F+1)}
                                     {\left[(2F+3)(2F+2)(2F+1)2F(2F-1)\right]^{1/2}} \nonumber \\
        && \times (-1)^{I+J+F}\left\{\begin{array}{rrr}J&2&J\\F&I&F\end{array}\right\}
              \left(\begin{array}{rrr}J&2&J\\-J&0&J\end{array}\right)^{-1}X \ ,
\label{eq:lqss}
\end{eqnarray}
where $A_{\mbox{\tiny{DC}}}$ is the electric field gradient,
$\Theta(\gamma,J)$ the quadrupole moment, and $X$ contains the
angular factors resulting from the rotation of the quadrupole field
frame to the quantization axis \cite{Ita00}. The quadrupole moment
of the $7s\,^2S_{1/2}$ ground state is zero, those of the $6d\,^2D_{3/2}$
and $6d\,^2D_{5/2}$ levels \cite{Sah07} are listed in Table \ref{table:pols}.
There are three special cases in which the first-order effect also vanishes
for particular hyperfine states of the $6d\,^2D_J$ levels:
\begin{enumerate}
\item[($i$)] $F=0$ levels have no quadrupole moment;
                   this applies to the $^{223,227,229}$Ra$^+$ cases.
\item[($ii$)] When $F=2$, $I=3/2$, $J=3/2$, the $6j$-symbol in
                    Eq. (\ref{eq:lqss}) is zero. This set of quantum numbers is available
                    in $^{223,227}$Ra$^+$, however, there is no improvement over the
                    previous case ($i$). All other shifts and associated uncertainties
                    were calculated to be equal to, or larger than, their counterparts
                    in the $F=0$ case. Therefore, these transitions have not been
                    included in Tables \ref{table:sensitivities} and
                    \ref{table:uncertainties}.
\item[($iii$)] For $F=3$, $m_F=\pm 2$ the shift vanishes because
                      of the factor $F(F+1)-3m_F^2$ in Eq. (\ref{eq:lqss});
                      this applies to the $^{225}$Ra$^+$(1) case.
\end{enumerate}
The transitions in $^{226}$Ra$^+$ and $^{225}$Ra$^+$(2) do suffer from
a linear quadrupole shift. These are given in Tables \ref{table:sensitivities}
and \ref{table:uncertainties}. To estimate the size of these shifts and their
uncertainties, we assumed that in the trap a typical static stray electric field
gradient $A_{\mbox{\tiny{DC}}} \simeq 10^3$ V/cm$^2$ is present due to
patch potentials. We assume that the angular factor $X$ is of order 1.
Since the orientation of the stray field is unknown, we take the full shift
as an estimate of the uncertainty. The effects of the much larger RF
trapping fields average out over the interrogation period.

However, the transitions that are free from the linear effect do suffer from a
second-order, quadratic quadrupole Stark shift. This contribution is significant
because now the effects from the RF trap potential do not average out. This RF
potential gives rise to a typical rms field gradient $A_{\mbox{\tiny{AC}}} = 10^4$
V/cm$^2$. To estimate the size of the shift we assume that the magnetic-field
orientation and the $z$-axis of the quadrupole trap field coincide. Taking only
couplings to hyperfine partners into account results in 
\begin{eqnarray}
   h\Delta\nu_{\mbox{\tiny{QQS}}}\left(\gamma,I,J,F,m_F\right)
   &=&4A^2_{\mbox{\tiny{AC}}}\Theta(\gamma,J)^2\sum_{F'}
   \frac{(2F+1)(2F'+1)}{E-E^\prime} \nonumber \\
   && \times \left(\begin{array}{rrr}F'&2&F\\
   -m_F&0&m_F\end{array}\right)^2\left\{\begin{array}{rrr}J&F'&I\\
   F&J&2\end{array}\right\}^2
\left(\begin{array}{rrr}J&2&J\\-J&0&J\end{array}\right)^{-2} \ .
\end{eqnarray}
It should be feasible to achieve an overall 10\% accuracy in the
determination of this shift, which is the uncertainty quoted in
Table \ref{table:uncertainties}.

It can be seen in Table \ref{table:uncertainties} that, similar to other clocks, the
linear quadrupole shift is by far the largest shift in Ra$^+$. In $^{199}$Hg$^+$
it was canceled by means of an averaging scheme \cite{Ita00,Ros08,Osk05},
which brought down the uncertainty level to the $10^{-17}$ level. An alternative
was presented more recently for $^{88}$Sr$^+$ in Ref. \cite{Dub05}, where it
is projected that the uncertainty caused by the electric quadrupole shift can be
reduced to the $10^{-18}$ level.

\section{Discussion and Conclusions}
Tables \ref{table:sensitivities} and \ref{table:uncertainties} contain the quantitative
results of our studies. Table \ref{table:sensitivities} lists the sensitivities of the isotopes
under study to the external fields. Also in Table \ref{table:sensitivities} the sensitivities
of three other ion clocks that are based on an electric quadrupole transition are shown
for comparison. In Table \ref{table:uncertainties}, the Ra$^+$ sensitivities have been
combined with typical values (and uncertainties) for the required and spurious external
fields to quantify the resulting shifts and the fractional frequency uncertainties
$\delta\nu/\nu$, where $\nu$ is the transition frequency and $\delta\nu$ the uncertainty
in the total shift. In the top half of the Table the different shifts  are given in mHz, with
the corresponding uncertainty between brackets.

The transitions $^{225}$Ra$^+(1)$, $^{226}$Ra$^+(1)$, and $^{226}$Ra$^+(2)$
suffer from the linear Zeeman (LZ) shift, which therefore has to be controlled to the
desired level of accuracy. The transitions $^{225}$Ra$^+(2)$, $^{226}$Ra$^+(1)$,
and $^{226}$Ra$^+(2)$ suffer from a linear quadrupole Stark (QS) shift of the order
6-24 Hz, which has to be cancelled in order for these cases to be competitive. As
mentioned, an  averaging scheme was implemented for $^{199}$Hg$^+$, a system
comparable to $^{225}$Ra$^+$(2), and $10^{-17}$ levels have been achieved
\cite{Ros08,Osk05}.
With an alternative averaging scheme, it appears feasible to reduce the QS shift
experimentally to the $10^{-18}$ level in $^{88}$Sr$^+$ \cite{Dub05}, a system
comparable to $^{226}$Ra$^+$. The transitions in $^{223}$Ra$^+$, $^{227}$Ra$^+$,
and $^{229}$Ra$^+$ are insensitive to both the LZ and the linear QS shifts, which
is in principle a clear experimental advantage. The quadratic QS shifts are only of
the order of 1 mHz.  $^{227}$Ra$^+$ is overall slightly better than $^{223}$Ra$^+$,
while $^{229}$Ra$^+$ is worse, because it has a relatively large quadratic Zeeman
(QZ) shift. As discussed, of these three, only $^{223}$Ra can be obtained from a source.

Provided that the LZ and linear QS shifts can be cancelled in $^{225,226}$Ra$^+$,
the largest remaining shift is caused by the BB radiation. It is of order 0.2 Hz in all
the isotopes. As in the case of $^{199}$Hg$^+$, this shift can be rendered negligible
by cooling down the system, albeit at the cost of a more complicated experimental
setup. For that reason the BB shift is given for two temperatures, namely for room
temperature (293 K) and for liquid-nitrogen temperature (77 K). The combination
of these two options with the possibility of averaging away the QS shift (indicated
by ``no QS'' in Table \ref{table:uncertainties}) give us in total four different results
for four sets of experimental choices, as shown in the bottom half of Table
\ref{table:uncertainties}. In the calculation of these uncertainties in the case ``no
QS,'' we have assumed that the LZ shift and the linear QS shift can be averaged
out experimentally to negligible values. The actual obtainable accuracies in these
cases depend on experimental details, but, as discussed, it appears realistic to
aim for accuracy levels of a few times $10^{-18}$.

We conclude that in particular the isotopes $^{223,225,226}$Ra$^+$ are promising
clock candidates with projected sensitivities that are all below the $10^{-17}$ level.
The actual experimental feasibility of the scenarios discussed above remains
to be demonstrated, of course. $^{223}$Ra$^+$ stands out as an attractive
simple candidate, without LZ and linear QS shifts, providing a compact, robust,
and low-cost atomic clock.

\section{Summary}
In summary, a theoretical analysis of the possible performance of a radium single-ion
optical clock was presented. It was shown that transitions in several readily available
Ra$^+$ isotopes are excellent candidates for alternative optical frequency standards.
The advantages of a heavy single ion that can be directly laser-cooled and interrogated
with off-the-shelf available semiconductor lasers are clear for many applications in
which costs and system size and stability are of importance. Furthermore, Ra$^{+}$
is an excellent laboratory for the search for variation of fundamental constants,  where
it ranks among the most sensitive candidates.

\section{Acknowledgments}
We thank N. Fortson, R. Hoekstra, and B. K. Sahoo for discussions, and W. Itano for
a helpful communication. This research was supported by the Dutch Stichting voor
Fundamenteel Onderzoek der Materie (FOM) under Programmes 114 and 125 and
FOM projectruimte 06PR2499. O. O. V. acknowledges funding from  the NWO
Toptalent program.

\bibliography{clockpaper}

\end{document}